\documentclass[elsart,usenatbib]{article}

\usepackage{psfig}
\usepackage{amsmath}
\usepackage{natbib}
\usepackage{graphicx}
\begin{document}

\title {The
iron $K_\alpha$ lines as a tool
for magnetic field estimations in non-flat accretion flows}
\author{A.F.Zakharov$^{1,2,3}$,
 Z.~Ma$^{1}$, Y.~Bao$^{4}$
\\
$^1$ National Astronomical Observatories, \\
Chinese Academy of Sciences, Beijing, 100012, China\\ $^2$ Russian
Scientific Centre --\\
Institute of Theoretical and Experimental Physics, 117259, Moscow,\\
$^3$
Astro Space Centre of Lebedev Physics Institute, Moscow\\
$^4$ Department of Mechanics, Zhongshan University,\\ Guangzhou,
510275, China }

\maketitle

\begin{abstract}

 Observations of AGNs and microquasars by ASCA, RXTE, Chandra
and XMM-Newton indicate the existence of broad X-ray emission
lines of  ionized heavy elements in their spectra. Such spectral
lines were discovered also in X-ray spectra of neutron stars and
X-ray afterglows of GRBs. Recently, \cite{ZKLR02} described a
procedure to estimate an upper limit of the magnetic fields in
regions from which X-ray photons are emitted. The authors
simulated typical profiles of the iron $K_\alpha$ line in the
presence of magnetic field and compared them with observational
data in the framework of the widely accepted accretion disk model.
Here we further consider typical Zeeman splitting in the framework
of a model of non-flat accretion flows, which is a generalization
of previous consideration into non-equatorial plane motion of
particles emitting X-ray photons.
 Using perspective facilities of space borne instruments  (e.g.
Constellation-X mission) a better resolution of the blue peak
structure of iron $K_\alpha$ line will allow to evaluate the
magnetic fields with higher accuracy.

\end{abstract}


\section{Introduction}

   Recent ASCA, RXTE, Chandra and XMM-Newton observations
of Seyfert~I galaxies have demonstrated the existence of the broad
iron $K_\alpha$ line (6.4~keV) in their spectra along with a
number of other weaker lines (Ne~X, Si~XIII, XIV, S~XIV-XVI,
Ar~XVII, XVIII, Ca~XIX, etc.) (see, for example,
\citealt{fabian1,tanaka1,nandra1,nandra2,malizia,sambruna,
yacoob4,ogle1}).

     For some cases when the spectral resolution is good enough,
the emission spectral line demonstrates the typical two-peak
profile with a  high "blue" peak and a low "red" peak while a long
"red" wing drops gradually to the background level
(\citealt{tanaka1,yaqoob2}, see also \citealt{Reynolds03} and
references therein). The Doppler line width corresponds to a very
high velocity of  matter.\footnote{Note that the detected line
shape differs essentially from the Doppler one.} E.g., the maximum
velocity is about $v \approx 80000 - 100000$~km/s for the galaxy
MCG--6--30--15 \citep{tanaka1,Fabian02} and $v \approx 48000$~km/s
for \mbox{MCG--5--23--16} \citep{krolik1}. For both galaxies line
profiles are known rather well. \cite{Fabian02} analyzed results
of long-time observations of MCG-6-30-15  using {\it XMM-Newton}
and {\it BeppoSAX}. They confirmed in general the qualitative
conclusions about the features of the Fe $K_\alpha$ line, which
were discovered by ASCA satellite. \citet{yaqoob02} discussed the
essential importance of ASCA calibrations and the reliability of
obtained results. \cite{Lee02} compared data among ASCA, RXTE and
Chandra  for the MCG-6-30-15. \cite{Iwa99,Lee99,Shih02} analyzed
in detail the variabilities in continuum and in Fe $K_\alpha$ line
for MCG-6-30-15 galaxy.

     The phenomena of the broad emission lines are supposed
to be related with accreting matter around black holes.
\cite{Wilms01,Ball01,Mart02} proposed physical models of accretion
discs for  MCG-6-30-15 and showed their influence on the Fe
$K_\alpha$ line shape. \cite{boller01} found the features of the
spectral line near  7~keV in Seyfert galaxies with data from {\it
XMM-Newton} satellite. \cite{yaqoob01a} presented results of
Chandra HETG observations of Seyfert~I galaxies. ~\cite{Qing01}
discussed a possible identification of binary massive black holes
with the analysis of Fe $K_\alpha$ profiles. \cite{Ball02} used
the data of X-ray observations to estimate an abundance of the
iron. \cite{Popov01,Popov02} discussed an influence of
microlensing on the distortion of spectral lines (including Fe
$K_\alpha$ line) that can be significant in some cases, optical
depth for microlensing in X-ray band was evaluated by
\cite{Zakharov04}. \cite{matt02} analyzed an influence of Compton
effect on emitted and reflected spectra of the Fe $K_\alpha$
profiles. In addition, \cite{fabian99a} presented a possible
scenario for evolution of such supermassive black holes.
\cite{moral01} proposed a procedure to estimate the masses of
supermassive black holes.

     General status of black holes was described in
a number of papers (see, e.g. \citealt{Liang98} and references
therein, \citealt{Zak00,FN01,Cherep03}). Since the matter motions
indicate very high rotational velocities, one can assume the
$K_\alpha$ line emission arises in the inner regions of accretion
discs at distances $\sim (1\div 3)~r_g$ from the black holes. Let
us recall that the innermost stable circular orbit for
non-rotational black hole (which has the Schwarzschild metric) is
located at the distance of $3\,r_g$ from the black hole
singularity. Therefore, a rotation of black hole could be the most
essential factor. A possibility to observe the matter motion in so
strong gravitational fields could give a chance not only to check
general relativity predictions and simulate physical conditions in
accretion discs, but investigate also observational manifestations
of such astrophysical phenomena like jets
\citep{romanova1,romanova2}, some instabilities like Rossby waves
\citep{Love99} and gravitational radiation.

     Observations and theoretical interpretations of broad
X-ray lines (particularly, the iron $K_\alpha$ line) in AGNs are
actively discussed in a number of papers
\citep{yaqoob1,wanders,sulentic1,sulentic2,paul,Bia02,Tur02,
Lev02a}. The results of numerical simulations are also presented
in the framework of different physical assumptions on the origin
of the broad emissive iron $K_\alpha$ line in the nuclei of
Seyfert galaxies
\citep{Matt92a,Matt92b,bromley,pariev2,pariev1,cui,bromley2,pariev3,
ma02a,Ma02,karas01}. The results of Fe $K_\alpha$ line
observations and their possible interpretation are summarized by
\citet{fabian2}.

According to the standard interpretation these $K_\alpha$ lines
are formed due to the  cold thin and optically thick accretion
disk illumination by hot clouds \citep{Fabian89,Laor91}, however
another geometry for regions of hot and cold clouds located near
black holes is not excluded. For example,
\cite{Hartnoll00,Hartnoll01,Hartnoll02,Blackman02} considered a
more complicated structure of accretion disks including warps,
clumps and spirals. \cite{Karas00} investigated a possibility to
explain Fe $K\alpha$ line with the model that the innermost part
of a disk is disrupted owing to disk instabilities and forms cold
clouds which move not exactly in the equatorial plane, but they
form a layer (or shell) covering a significant part of sky from
the point of view of central X-ray source (actually, that is a
detailed analysis of ideas suggested by \cite{Collin96}). Other
features of such a model were discussed by \cite{Malzac01}. An
influence of warps on X-ray emission line shapes were investigated
recently by \cite{Cadez03} analyzing photon geodesics in the
Schwarzschild black hole metric.

     Broad spectral lines are considered to be formed by
radiation  emitted in the vicinity of black holes.  If there are
strong magnetic fields near black holes these lines are split by
the field into several components. Such lines have been found in
microquasars, GRBs and other similar objects
\citep{Balu99,grein99,mira00,Lazz01,Mart02a,Mira02a,
Miller02,zaman02}.

     To obtain an estimation of the magnetic field we simulate
the formation of the line profile for different values of magnetic
field in the framework of the simple model of  non-flat accretion
flows assuming that emitting particles move along orbits with
constant radial coordinates, but not exactly in the equatorial
plane. Earlier, \cite{ZKLR02} analyzed an influence of magnetic
field on a distortion of $K_\alpha$ line considering equatorial
circular motion of emitting region of the Fe $K_\alpha$ line
radiation.\footnote{Recently, \cite{Loeb03} has analyzed Zeeman
splitting for X-ray absorbtion lines in the X-ray spectrum of the
bursting neutron star EXO 0748-676.} Here we will use the simple
model of a non-flat accretion flow \citep{ma02a,Ma02}
 to analyze
the non-equatorial plane motion of particles emitting X-band
photons. Actually, we will use a generalization of the previous
annulus model described earlier by \cite{zak_rep1}.
 As a result we find the minimal $H$ value of magnetic field at which the
distortion of the line profile becomes significant. Here we do not
use an approach, which is based on numerical simulations of
trajectories of the photons emitted by annuli moving along a
circular geodesics near black hole, described earlier by
\cite{zakharov6,zakharov1,zakharov5,zak_rep1}. In this paper  we
generalize previous considerations for the the simple model of
non-flat accretion flow.

\section{Magnetic fields in accretion discs}

     Magnetic fields play a key role in dynamics of accretion
discs and jet formation. \cite{Bis74,Bis76} considered a scenario
to generate super strong magnetic fields near black holes.
According to their results magnetic fields near the marginally
stable orbit could be about $H \sim 10^{10} - 10^{11}$~G. However,
if we use a model of the Poynting -- Robertson magnetic field
generation then only small magnetic fields are generated
\citep{BKLB02}.
 \citet{Kard95,Kard00,Kard01a,Kard01} has shown that
the strength of the magnetic fields near super massive black holes
can reach the values of $H_{max} \approx 2.3\cdot 10^{10}
M_9^{-1}$~G due to the virial theorem\footnote{Recall that
equipartition value of magnetic field is only $\sim 10^4$~G.}, and
considered a generation of synchrotron radiation, acceleration of
$e^{+/-}$ pairs and cosmic rays in magnetospheres of super massive
black holes  at such high fields. It is the magnetic field that
plays a key role in these models. Below, based on the analysis of
iron $K_\alpha$ line profile in the presence of a strong magnetic
field, we describe how to detect the field itself or at least
obtain an upper limit of the magnetic field.

One of the basic problems in understanding the physics of quasars
and microquasars is the "central engine" in these systems, in
particular, a physical mechanism to accelerate charged particles
and generate energetic electromagnetic radiation near black holes.
The construction of such "central engine" without magnetic fields
could  hardly ever be possible. On the other hand, magnetic fields
make it possible to extract energy from rotational black holes via
Penrose process and Blandford -- Znajek mechanism, as it was shown
in MHD simulations by \cite{Koide02,Koide02a}. The Blandford --
Znajek process could provide huge energy release in AGNs (for
example, for MCG-6-30-15) and microquasars when the magnetic field
is strong enough \citep{Wilms01}.

    Physical aspects of generation and evolution
of magnetic fields were considered in a set
of reviews (e.g. \cite{Ass87,Giov01}).
A number of papers conclude that in the vicinity of
the marginally stable orbit the magnetic fields could be
high enough \citep{Bis74,Bis76,Krolik99}.
     \cite{Agol99} considered the influence of magnetic
fields on the accretion rate near the marginally stable orbit and
hence on the disc structure. They found the appropriate changes of
the emitting spectrum and solitary spectral lines. \cite{vietri98}
investigated the instabilities of accretion discs in the case when
the magnetic fields play an important role.
\cite{Li02a,Li02b,WLM03} analyzed an influence of magnetic field
on accretion disk structure and its emissivity through the
magnetic coupling of a rotating black hole with its surrounding
accretion disk.

Magnetic field could play a key role in Fe $K\alpha$ line
emission, since coronae around  accretion disks could be magnetic
reservoirs of energy to provide a high energy radiation
\citep{Merloni01} or magnetic flares could help to understand an
origin of narrow Fe $K\alpha$ lines and their temporal dependences
\citep{Nayakshin01,Nayakshin02}. \cite{Collin03} calculated X-ray
spectrum for the flare model  and pointed out some signatures of
the model to distinguish it from the well-known lamppost model
where it is assumed that an X-ray source illuminates the inner
part of accretion disk in a relatively steady way.

\section{Influence of a magnetic field on the distortion
         of the iron $K_\alpha$ line profile}

    The magnetic pressure at the inner edges of the accretion
discs and its correspondence with the black hole spin parameter
$a$ in the framework of disc accretion models is discussed
by~\cite{krolik01}. However, the numerical value of magnetic field
is determined there from a model-dependent procedure, in which a
number of parameters cannot be found explicitly from observations.

    Here we consider the influence of  magnetic field on
the iron $K_\alpha$ line profile \footnote{We can also consider
X-ray lines of other elements emitted by the area of accretion
disc close to the marginally stable orbit; further
we talk only
about iron $K_\alpha$ line for brevity.} and show how one can
determine the value of the magnetic field strength or at least
an upper limit.

    The profile of a monochromatic line
\citep{zak_rep1,zak_rep2,zak_rept} depends on the angular momentum
of a black hole, the inclination angle of observer, the value of
the radial coordinate if the emitting region represents an
infinitesimal ring (or two radial coordinates for outer and inner
bounds of a wide disc). The influence of accretion disc model on
the profile of spectral line was discussed by
\cite{zak_rep3,Zakharov_Repin_Ch03}.

    We assume that the emitting region is located in
the area of a strong quasi-static magnetic field. This field
causes line splitting due to the standard Zeeman effect. There are
three characteristic frequencies of the split line that arise in
the emission \citep{Blokh63,Dir58,Mess99}. The energy of central
component $E_0$ remains unchanged, whereas two extra components
are shifted by $\pm \dfrac{eH}{2mc}=\pm \mu_B H$, where
$\mu_B=\dfrac{e \hbar}{2m_{\rm e}c}=9.273\cdot 10^{-21}$~erg/G is
the Bohr magneton. Therefore, in the presence of a magnetic field
we have three energy levels: $E_0-\mu_B H,~ E_0$ and $E_0+\mu_B
H$. For the iron $K_\alpha$ line they are as follows:
$E_0=6.4\left(1 - \dfrac{0.58}{6.4} \cdot \dfrac{H}{10^{11}\,{\rm
G}} \right) $ keV, $E_0=6.4$~keV and $E_0=6.4\left(1  +
\dfrac{0.58}{6.4} \cdot \dfrac{H}{10^{11}\,{\rm G}}\right) $ keV.

\cite{Loeb03} pointed out that for a strong field, there is also a
net blueshift of the centroid of the transition line component
which is quadratic in $B$. For hydrogen-like ions
\cite{Jenkins39,Schiff39,Preston70} give

\begin{eqnarray}
(\Delta E) _{\rm shift} \sim \frac{e^2
a_0^2}{8Z^2m_ec^2}n^4(1+M_l)B^2=
 \nonumber \\
 = 9.2\times 10^{-4}
{\rm eV}
\left(\frac{Z}{26}\right)^{-2}n^4(1+M_l^2)\left(\frac{B}{10^9 {\rm
G}}\right),
\end{eqnarray}
where $n$ and $M_L$ are the principal and orbital quantum numbers
of the upper state, $a_0$ is the Bohr radius, and $Z$ is the
nuclear charge (=26 for Fe).

   Let us discuss how the line profile changes when photons
are emitted in the co-moving frame with energy $E_0 (1+\epsilon)$,
but not with $E_0$. In that case the line profile can be obtained
from the original one by $1+\epsilon$ times stretching along the
energy axis, the component with $E_0 (1-\epsilon)$ energy should
be $(1-\epsilon)$ times stretched, respectively. The intensities
of different Zeeman components are approximately equal
\citep{Fock76}, each of which depends on the direction of the
quantum escape with respect to the direction of the magnetic field
\citep{BLP89}. However, we neglect this weak dependence
(undoubtedly, the dependence can be counted and, as a result, some
details in the spectrum profile can be slightly changed, but the
qualitative picture, which we discuss, remains unchanged). As a
consequence, the composite line profile can be found by summation
the initial line with energy $E_0$ and two other profiles,
obtained by stretching this line along the $x$-axis in
$(1+\epsilon)$ and $(1-\epsilon)$ times correspondingly.

     Another indicator of the Zeeman effect is a significant
induction of the polarization of X-ray emission: the extra
lines possess a circular polarization (right and left,
respectively, when they are observed along the field direction)
whereas a linear polarization arises if the magnetic field is
perpendicular to the line of sight.\footnote{Note that another
possible polarization mechanisms in $\alpha$-disc were discussed
by~\citet{Saz02}.} Despite of the fact that
the measurements of polarization of X-ray emission have not
been carried out yet, such experiments can
be realized in the nearest future \citep{Cos01}.

     With increase of the magnetic field
the peak profile structure becomes apparent and can be distinctly
revealed, however, the field $H \sim 2\cdot 10^{11}$~G is rather
strong, so the classical linear expression for the Zeeman
splitting
\begin{equation}
     \epsilon=\frac{ \mu_B H}{E_0}
      \label{eq15a}
\end{equation}
should be modified. Nevertheless, we use Eq.(\ref{eq15a}) for any
value of the magnetic field, assuming that the qualitative picture
of peak splitting remains unchanged, whereas for $H = 2\cdot
10^{11}$~G the exact maximum positions may appear slightly
different. If the Zeeman energy splitting $\Delta E$ is of the
order of $E$, the line splitting due to magnetic fields is
described in a more complicated way. The discussion of this
phenomenon is not a point of this paper. Our aim is to pay
attention to the qualitative features of this effect.

Thus, besides magnetic field, the line profiles depend on the
accretion model as well as on the structure of emitting regions.
Problems of such kind may become actual with much better accuracy
of observational data in comparison with their current state.

\section{Non-flat accretion flows and  iron K$\alpha$ line shapes}

The relativistic generalization of Liouville's theorem was used to
calculate the spectral flux by many authors (e.g.
\citealt{thorne67,ames,gerlach}). Just after \cite{thorne74}
  finished the analysis of the time-averaged
structure of a thin, equatorial disk of material accreting onto a
black hole, \cite{cunningham} used Liouville's theorem to give a
prediction about the X-ray continuum from the disk.

Simulations of iron line profiles started from the paper by
\cite{Fabian89}. These calculations are based on assumptions about
geometrically thin, optically thick disks.
 These results were generalized by \cite{Laor91}
 to a Kerr black hole case. Many authors
(e.g. \citealt{bromley,dabrowski}) also used the thin-disk model
with Cunningham's approach: The solid angle was evaluated as a
function of both the emission radius $r$ and the frequency shift
$g$; The propagation of the line radiation was considered to emit
from the thin disk in the range from the innermost radius $r_{in}$
to the outermost radius $r_{out}$. Thus, the flux carried by a
bundle of photons from a whole disk needs the integration over
$r$. In this case, a single value of $\theta=90^\circ$ was fixed
in advance to simulate the thin-disk. Another approach to simulate
line profiles for flat accretion disk was proposed by
\cite{zakh91,zakharov1,zakharov5,zak_rep1,Zak02pr,zak_rep2,
zak_rept,Zak_rep02_xeus,Zak_rep02_Gamma,Zak_Rep03_Lom,Zak_rep03_aa,Zak_rep03_azh,Zak03_Sak,Zakharov_Repin_Pom04,zak_rep3}
which is based on qualitative analysis developed in previous
papers by \cite{zakh86,zakh89}.

In fact, different from the continuum, iron $K\alpha$ line is
believed to originate via fluorescence in the very inner part of a
disc (e.g. \citealt{tanaka1,iwasawa}), within which particles
exist in spherical orbits between the minimum and maximum
latitudes about the equatorial plane of the central black hole
\citep{wilkins,ma00} in the form of a hot torus \citep{chen} or a
shell (or a layer) formed by cold clouds which could be
illuminated by hot clouds \citep{Karas00,Malzac01}. Therefore, the
mechanism of the Fe line emissions should be re-considered with
the non-disk formulation, which is connected with several
parameters in both sets of coordinates, such as the spin of the
black hole, particles' constants of motion, photons's impact
parameters, the thickness and the radial position of the shell,
the polar angle of the shell, etc. In our work, as done by
previous authors with the thin-disk model
\citep{gerlach,Laor91,dabrowski}, we make following assumptions:
(1) The emitting "shell" is geometrically thin; i.e., at radius
$r$ its thickness $\delta r$ is always much less than $r$. This
permits us to treat particles as a thin-shelled ensemble at $r$
surrounding a BH. (2) The emission is isotropic and the emitted
fluorescent Fe K$\alpha$ line from particles can be described by a
$\delta$-function in frequency, which gives each emitted photon an
energy of 6.4~keV. That is, particles are monochromatic. (3)
Photons emitted by shell particles are homogeneous and are free to
reach the observer.

For a given $a$, there are sets of three constants of motion at
one radius $r$. That is, a thin shell means a collection of sets
of three constants of motion, $q$, $p$, and $\varepsilon^{2}$.
Considering the monotonous relations between $p$ and $q$ (or
$\varepsilon^{2}$), the number of the sets can be simply
represented by $q$. Therefore, different from the fact that the
continuum is an integration over $r$ in disk models, the line flux
should be a superimposition of all possible individual emissions
with every $q$ at $r$.

With the thin-disk model, previous authors (e.g.,
\citealt{Laor91}) considered the emitted intensity $I_{e}(r,\
\nu)$ as a created parameter $I_{e}(r,\
\nu)=\delta(\nu-\nu_{e})J(r)$, in which $J(r)$ is defined as "the
line-emissivity law" with different artificial forms versus the
radius of emission $r$ only; $\nu_{e}$ is the rest frequency of
the emission. Fortunately, a more realistic form of $J$ was
deduced by \cite{george}, in which $J$ is expressed as a function
of $r$, $\nu$ and $\theta$. However, Liouville's equation contains
not only the invariant photon four-momentum, but the four-velocity
\& four-coordinate components of particles as well.  Let $I$ (ergs
s$^{-1}$ cm$^{-2}$ sr$^{-1}$ eV$^{-1}$) be the specific intensity
and $N$ (cm$^{-3}$ dyn$^{-3}$ s$^{-3}$) the photon distribution
function. The relationship between $I$ and $N$ is
(\cite{thorne67}): $I=\dfrac{1}{2}\times2hN\nu_e^3$, where $h$ is
Planck constant, $\nu_e=\dfrac{6.4\mathrm{~keV}}{h}$ is the
K$\alpha$ frequency of an iron atom with 4-coordinate $x^{\mu}$
and 4-velocity $u^{\mu}$. Coefficient ${1}/{2}$ means only half of
photons emitted outwards and 2 indicates that there exists both
states of the photon quantum per phase-space. The expression of
$N$ is (\cite{gerlach})
\begin{equation}
N(x^{\mu},\ u^{\mu},\ s^{\mu})=C(x^{\mu})\ \cdot\ \delta
(u^{\mu}s_{\mu}-h\nu_e)
\end{equation}
where $C(x^{\mu})$ (cm$^{-3}$) is photon's number density.
According to the last assumption, $C=1$.

The dimensionless relative flux versus the shift
$\dfrac{\nu}{\nu_e}$ depends on three parameters: the black hole
spin $a$ , the radial position $r$ of emitters, and the
inclination angle ${\rm Obs}$ of the observer. The flux expression
is (c.f., e.g., \cite{Laor91,bromley})
\begin{equation}
F=\frac{r_0^2}{h\nu_e}\ F_{\rm line}=\frac{r_0^2}{h\nu_e}\
\sum_q\int d\nu d\Omega\cdot I\
\cdot\left(\frac{\nu}{\nu_e}\right)^3\cdot\mathrm{\cos}\alpha
\end{equation}
in which the integration over the element of the solid angle
$d\Omega$ covers the image of the spherical ring in the observer's
sky plane; the solid angle is expressed by impact parameters
$\alpha$, $\beta$ of the observer which are related to the
constants of motion; $E=h\nu$; $\dfrac{\nu}{\nu_e}$ is the general
relativistic frequency shift;
the sum $\sum$ is to all values of $q$, which reflects that the
flux is contributed by all photons emitted from all particles (see
\citealt{Ma02} for details).

This model could be interpreted as a simplified version of some
distribution of clouds in the "quasi-spherical" accretion model
developed earlier by \cite{Celotti92,Collin96}  to evaluate
optical/UV/soft X-ray emission of AGNs. We use such a distribution
of clouds to calculate the Fe $K\alpha$ line shapes in presence of
a strong magnetic field which could play a significant role in
such models \citep{Celotti92}.

In numerical simulations of photon geodesics we used their
analytical analysis to reduce numerical errors.
    A qualitative analysis of the geodesic equations
showed that types of photon motion can drastically vary with small
changes of chosen geodesic parameters \citep{zakh86,zakh89}. In
our approach we use results of this analysis and numerical
calculations of photon geodesics \citep{zakh91}.

\section{Simulation results}

We have calculated spectral line shapes for different parameters
of the model. Below we briefly describe results of these
calculations.

\begin{figure*}[!tbh]
\begin{center}
\includegraphics[width=0.98\textwidth]{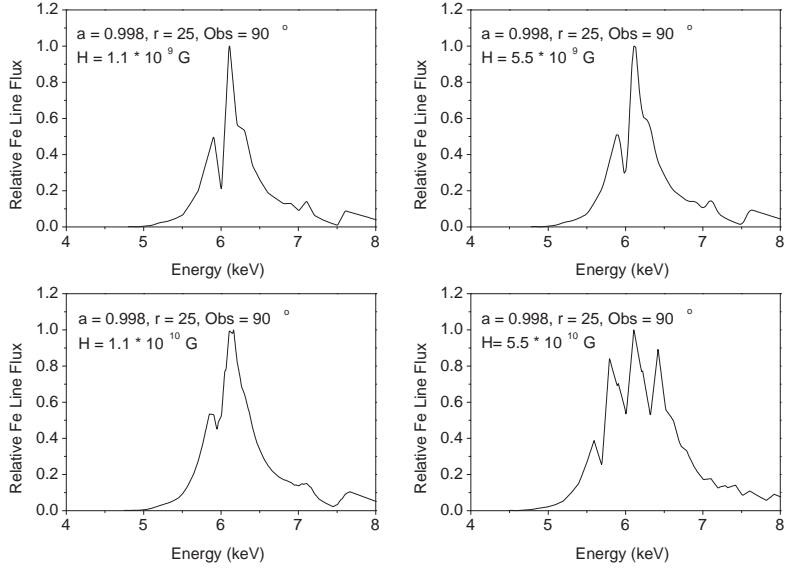}
\end{center}
  \vspace{-7mm}
  \caption{Spectral line shapes  for $a=0.998$, $r=25$, the inclination angle of
  observer (Obs$=90^\circ$),
           for different magnetic fields $H=1.1 \times 10^9$ G, $5.5 \times
           10^9$G, $1.1 \times 10^{10}$ G, $5.5 \times  10^{10}$G.}
  \label{figure2}
\end{figure*}

Fig. \ref{figure2} shows a series of spectral line profiles for
$a=0.998$,
 $r=25$ and the inclination angle of observer
Obs$=90^\circ$ (it means that an observer is located in the
equatorial plane)
    for magnetic fields $H=1.1 \times 10^9$ G, $5.5 \times
           10^9$G, $1.1 \times 10^{10}$ G, $5.5 \times  10^{10}$G respectively.
           Corresponding parameter values are indicated in the
           left top angle of each panel. One could see from these
           panels that magnetic field  $H= 5.5 \times 10^9$G does
           not distort significantly the spectral line profiles,
           but  $1.1 \times 10^{10}$G gives a significant
           broadening the peaks of the profile and as a result the shortest red
           peak may be not distinguishable from observational
           point of view. The spectral line profile for $H=5.5 \times  10^{10}$G
demonstrates a significant difference from the spectral line
profile with respectively "low" magnetic field like in first panel
$H=1.1 \times 10^9$ G, because there is an evident three peaked
structure of the spectral line profile for  $H=5.5 \times
10^{10}$G.

\begin{figure*}[!tbh]
\begin{center}
\includegraphics[width=0.98\textwidth]{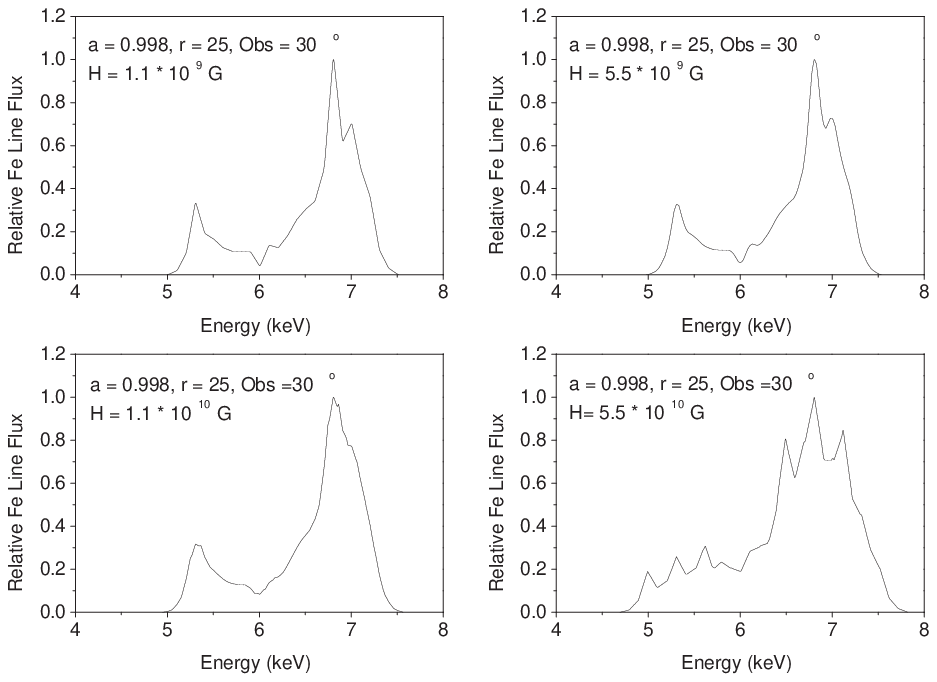}
\end{center}
  \vspace{-7mm}
  \caption{Spectral line shapes  for $a=0.998$, $r=40$, Obs$=50^\circ$,
           for different magnetic fields $H=1.1 \times 10^9$G, $5.5 \times
           10^9$G, $1.1 \times 10^{10}$G, $5.5 \times  10^{10}$G.
           }
  \label{figure3}
\end{figure*}
Fig. \ref{figure3} shows a series of spectral line profiles for
$a=0.998$,
 $r=40$ and
  Obs$=50^\circ$ and for magnetic fields  $H=1.1 \times 10^9$ G, $5.5 \times
           10^9$G, $1.1 \times 10^{10}$ G, $5.5 \times  10^{10}$G, respectively.
In this case
the magnetic field $H= 5.5 \times 10^9$G also does not distort
significantly the spectral line profiles,
           but  $H=1.1 \times 10^{10}$G gives a significant
           broadening the profile and as a result the shortest extra blue
 peak is disappeared and the initial structure of the the spectral line profile
with  high and low blue peaks and a red peak is changed to a
two-peaked structure with a red peak and a blue peak.  As in Fig.
\ref{figure2},  the spectral line profile for $5.5 \times
10^{10}$G demonstrates a significant difference from the spectral
line profile with respectively "low" magnetic field like in the
first panel $H=1.1 \times 10^9$ G. This result is general for our
calculations and applicable for all cases considered below.

\begin{figure*}[!tbh]
\begin{center}
\includegraphics[width=0.98\textwidth]{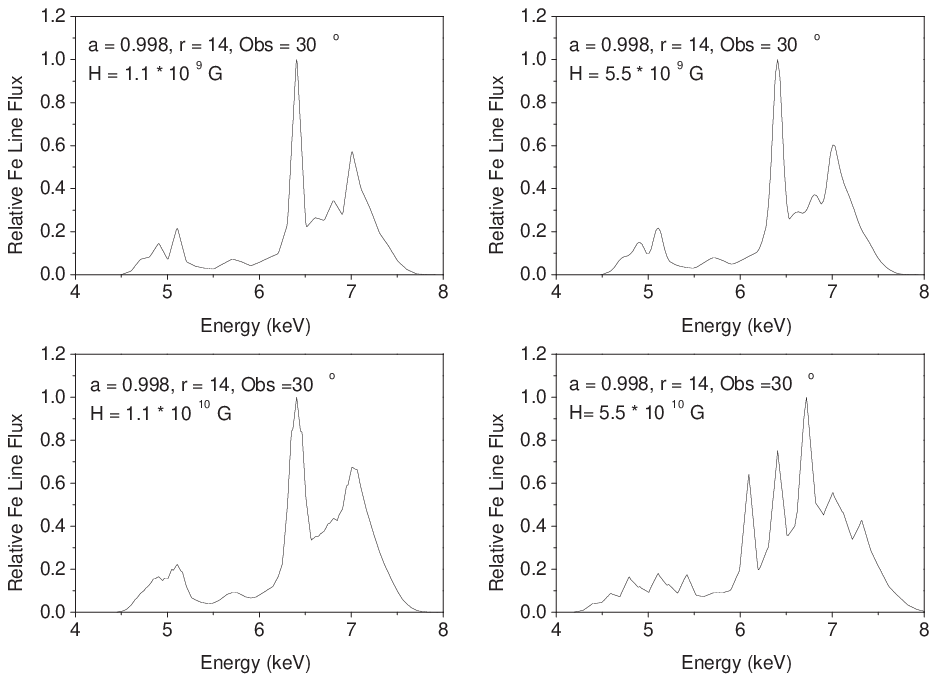}
\end{center}
  \vspace{-7mm}
  \caption{Spectral line shapes  for $a=0.998$, $r=14$,
  Obs$=30^\circ$,
           for different magnetic fields $H=1.1 \times 10^9$ G, $5.5 \times
           10^9$G, $1.1 \times 10^{10}$ G, $5.5 \times  10^{10}$G.
           }
  \label{figure3a}
\end{figure*}

Fig. \ref{figure3a} shows a series of spectral line profiles for
$a=0.998$,
 $r=14$
and  Obs$=30^\circ$ for magnetic fields  $H=1.1 \times 10^9$ G,
$5.5 \times
           10^9$G, $1.1 \times 10^{10}$ G, $5.5 \times  10^{10}$G,
           respectively.
           In this case the blue peak is higher than the red one
(such types of peaks were calculated also by \cite{Cadez03} for
the warped disks).
           As for previous case (Fig. \ref{figure2}) magnetic
field $H= 5.5 \times 10^9$G does not distort significantly the
spectral line profiles,
           but  $H=1.1 \times 10^{10}$G gives a significant
           broadening the peaks of the profile and as a result
sub-peaked structure between blue and red peaks is disappeared.

\begin{figure*}[!tbh]
\begin{center}
\includegraphics[width=0.98\textwidth]{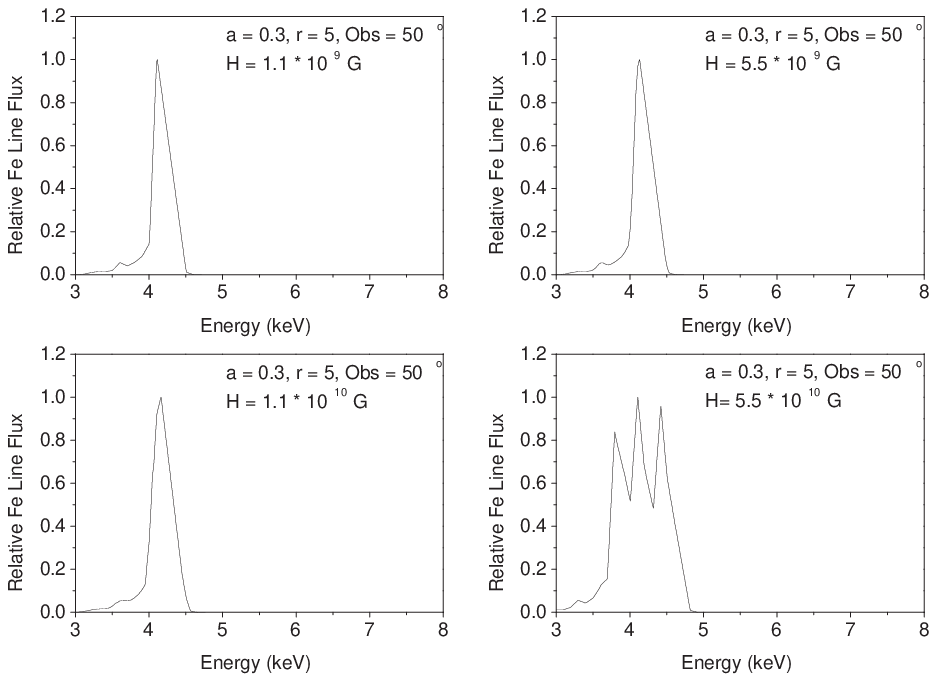}
\end{center}
  \vspace{-7mm}
  \caption{Spectral line shapes  for $a=0.3$, $r=5$,
  Obs$=50^\circ$,
           for different magnetic fields $H=1.1 \times 10^9$ G, $5.5 \times
           10^9$G, $1.1 \times 10^{10}$ G, $5.5 \times  10^{10}$G.}
  \label{figure4}
\end{figure*}

\begin{figure*}[!tbh]
\begin{center}
\includegraphics[width=0.98\textwidth]{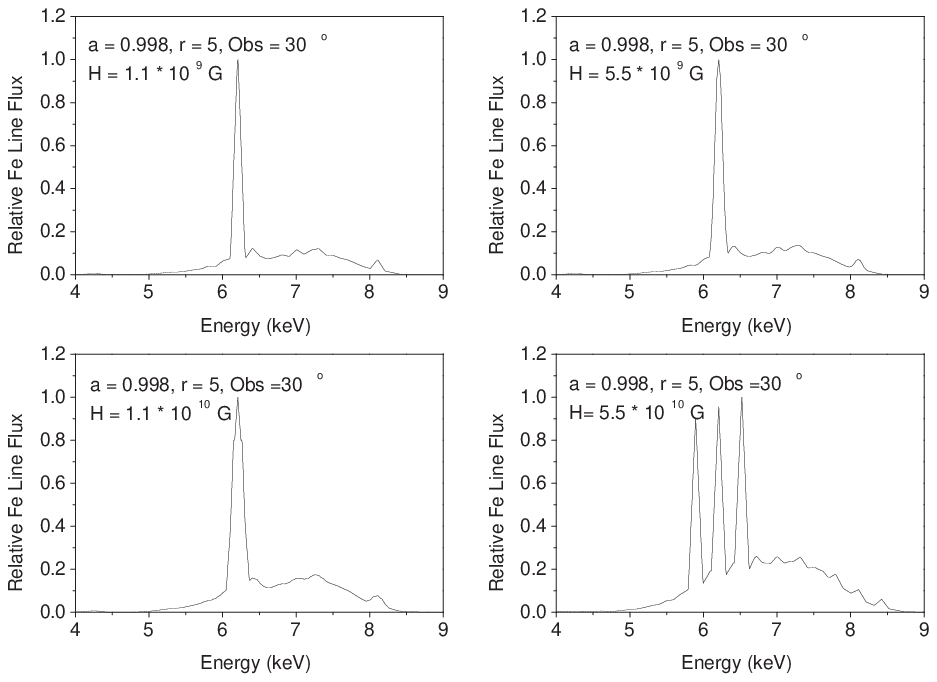}
\end{center}
  \vspace{-7mm}
  \caption{Spectral line shapes  for $a=0.998$, $r=5$, Obs$=30^\circ$
           for different magnetic fields $H=1.1 \times 10^9$ G, $5.5 \times
           10^9$G, $1.1 \times 10^{10}$ G, $5.5 \times  10^{10}$G.}
  \label{figure5}
\end{figure*}

\begin{figure*}[!tbh]
   \begin{center}
\includegraphics[width=0.98\textwidth]{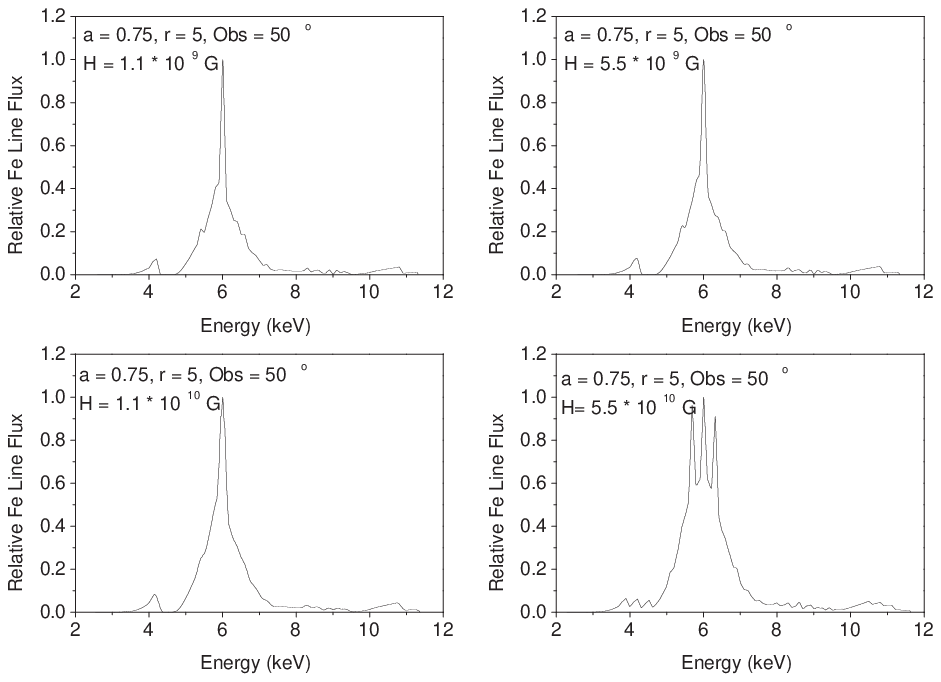}
\end{center}
  \caption{Spectral line shapes  for $a=0.75$, $r=5$,
  Obs$=50^\circ$,
           for different  magnetic fields $H=1.1 \times 10^9$ G, $5.5 \times
           10^9$G, $1.1 \times 10^{10}$ G, $5.5 \times  10^{10}$G.}
  \label{figure6}
\end{figure*}

The Zeeman split of one peaked spectral line profiles is presented
in Fig. \ref{figure4}-\ref{figure6}. Similar spectral line
profiles (without Zeeman splitting) were calculated in the
framework of a warped disk model by \cite{Cadez03} and for a cloud
model by \cite{Karas00}. Fig. \ref{figure4} shows that for
$a=0.3$, $r=5$ and Obs$=50^\circ$, there is only one single peak,
but there is
  no bump around this peak for low magnetic fields.
Fig. \ref{figure5} shows that for $a=0.998$, $r=5$ and
Obs$=30^\circ$, there is also only one single peak, but there is
  a broad bump around this peak for low magnetic fields (evidences
  for such a kind of bump was found by \cite{Wilms01} using data
  of XMM-EPIC observations of MCG-6-30-15\footnote{\cite{Wilms01} suggested that there
  are magnetic fields ($h \sim 10^4$\,G) in the Seyfert galaxy MCG-6-30-15
  and even there is magnetic extraction of energy because of Blandford -- Znajek
  effect, but   of course, these magnetic fields are too low to lead to a
  significant changes of the iron $K_\alpha$ line due to Zeeman splitting.}).
Fig. \ref{figure6} shows that for $a=0.75$, $r=5$ and
  Obs$=50^\circ$,  there is also only one single peak and there is
  a bump around this peak for low magnetic fields.
Figs. \ref{figure5},\ref{figure6} demonstrate two cases where
Zeeman splitting gives significant changes of spectral line
profiles, in which evident single peak structure for low magnetic
fields is changed into a three-peaked structure for high magnetic
fields $H \sim 5.5 \times 10^{10}$G.

Summarizing these results of calculations for considered examples,
we note that magnetic fields $H \sim 5 \times 10^{10}$G produce
significant changes of spectral line profiles, but $H \sim
10^{10}$ G could be responsible for essential broadening the
profile peaks. Therefore, in principle there is a possibility to
measure magnetic fields about $5 \times 10^{10}$G that could be
generated near Galactic Black Hole Candidates and probably near
black hole horizons in some AGNs.

\section{Summary and discussions}

It is known that Fe $K_\alpha$ lines are found not only in AGNs
and microquasars but also in X-ray afterglows of gamma-ray bursts
(GRBs) \citep{Lazz01}. A theoretical model for GRBs was suggested
recently by \cite{Van_Putten01,Van_Putten03}. In the framework of
this model a magnetized torus (shell) around rapidly rotating
black hole could be formed after black hole-neutron star
coalescence. In this case magnetic fields could be even much
higher than $10^{11}$~G. Therefore, an influence of magnetic
fields on spectral line profiles can be very significant and we
must take into account Zeeman splitting.

 Results of 3D
magnetohydrodynamical (MHD) simulations  demontrated that there
are non-equatorial and non-axisymmetric density patterns and some
configurations like tori or shells could be formed
\citep{Mineshige02}. Moreover, an analysis of instabilities of
accretion flows showed that warps, tilts and caustic surfaces
could arise not only in the equatorial plane \citep{Illarionov01}.
Observations also gave some indirect evidences for more
complicated accretion flows (than the standard thin accretion
flow) because there are some signatures for a precession and a
nutation  (for example, there is a significant precession of the
accretion disk for the SS433 binary system
\citep{Cher02}).\footnote{\cite{Shak72} predicted that if the
plane of an accretion disk is tilted relative to the orbital plane
of a binary system, the disk can precess.}

    It is evident that duplication (triplication)
of a blue peak could be caused not only by the influence of a
magnetic field (the Zeeman effect), but by a number of other
factors. For example, the line profile can have multiple peaks
when the emitting region represents multiple shells with different
radial coordinates (it is easy to conclude that two emitting rings
with finite widths separated by a gap, would yield a similar
effect). Actually, such an explanation was proposed by
\cite{Tur02} to fit the Fe $K\alpha$ shape in NGC 3516.
 Despite
of the fact that a multiple blue peak can be generated by many
causes (including the Zeeman effect as one of possible
explanation), the absence of the multiple peak can lead to a
estimation of an upper limit of the magnetic field.

    It is known that neutron stars (pulsars) could have
huge magnetic fields. So, it means that the effect discussed above
could appear in binary neutron star systems and in single neutron
stars as well \citep{Loeb03}. The quantitative description of such
systems, however, needs more detailed computations.

Similar to considerations presented in paper by \cite{ZKLR02},
analyzing Fe $K\alpha$ shapes for MCG-6-30-15 galaxy and using
ASCA data \citep{tanaka1} one could evaluate a magnetic field for
this case; namely a magnetic field should be less than $5 \times
10^{10}$~G and the estimate is independent on a character of
accretion flow. So, we could use the estimate for non-flat
accretion flows for MCG-6-30-15 Seyfert galaxy and generalize
conclusions by \cite{ZKLR02} for more general cases of accretion
flows.

As an extended work of the first paper by Zakharov et al. (2003),
the estimates of magnetic field may seem not very precise. But one
could mention that the rough estimates are caused by a noisy
observational data since as a matter of fact, present spaceborne
instrumentations vary greatly in their sensitivities and
resolutions and precisions of measurements are not very high to
have good estimates. Moreover, it would be difficult to reveal a
compromisable list quantitatively of possible limiting effects
restricted by different detections. However, \cite{Zakharov_Ma04}
are trying to focus on  specific ASCA observations of PKS 0637-752
in the range 1.3-24.8 keV \citep{Yaqoob98} and provide an
estimation of the minimum magnetic field detectable with the
satellite taking into account  a new X-ray emission hypothesis
\citep{Varshni99}, other than a fluorescent assumption. The
analysis is based on the laboratory measurements and
identification of iron line experiments in Lawrence Livermore
National Laboratory \citep{Brown02}.

    With further increase of observational facilities it
may become possible to improve the above estimation. The
Constellation-X launch suggested in the coming decade seems to
increase the precision of X-ray spectroscopy as many as
approximately 100 times with respect to the present day
measurements \citep{weaver1}. Therefore, there is a possibility in
principle that the upper limit of the magnetic field can also
greatly improved in the case when the emission of the X-ray line
arises in a sufficiently narrow region.

\section{Acknowledgements}
Authors are grateful to J.-X.~Wang for fruitful discussions.
Authors thank an anonymous referee for very useful remarks.

 This work was supported by the National Natural Science Foundation
of China, No.:10233050.

\end{document}